\documentclass{article}
\usepackage{authblk}
\usepackage{setspace}
\usepackage{blindtext}
\usepackage{indentfirst}
\usepackage{amsmath}

\begin{document}

\title{Exact Solutions for the Fluid Impulse for Incompressible and Compressible Flows}
\date{}
\author{M. Michalak and B. K. Shivamoggi\\ University of Central Florida \\ Orlando, Fl, 32816-1364 }
\affil{}
\maketitle

\begin{abstract}
In this paper, we consider the Kuz'min\cite{Kuzmin}-Oseledets\cite{Oseledets} formulations for the fluid impulse density for the incompressible and compressible barotropic Euler equations. We give exact solutions for the fluid impulse density and discuss their physical significance. The compressible barotropic exact solution describes a density wave riding on a steady Burgers type vortex flow with the fluid impulse velocity component in the wave-propagation direction growing monotonically with time.

\end{abstract}

\section{Introduction}
\par Impulse formulations of Euler (and Navier-Stokes) equations were initiated by Kuz'min \cite{Kuzmin} and Oseledets \cite{Oseledets} and have been found to be useful in dealing with topological properties of vortex lines. Russo and Smereka \cite{RussoSmereka} exhibited different impulse formulations upon using various gauge transformations. In the Kuz'min-Oseledets formulation, the fluid impulse velocity is given by

\begin{equation} 
\textbf{p}=\textbf{v} + \nabla \varphi.
\label{eq1}
\end{equation}
(\ref{eq1}) defines a gauge transformation with the scalar field $\varphi$ serving as the gauge. The geometric gauge, which is defined such that $\varphi$ satisfies,
\begin{equation}
\frac{D\varphi}{Dt} = P-\frac{1}{2} \left| \textbf{v} \right| ^2 ,
\label{eq2}
\end{equation}
transforms the equation of motion for ideal hydrodynamics,
\begin{equation}
\frac{D\textbf{v}}{Dt} = - \frac{\nabla P}{\rho},
\label{eq3}
\end{equation}
into

\begin{equation}
\frac{D\textbf{p}}{Dt} = -(\nabla \textbf{v})^T \cdot \textbf{p}
\label{eq4}
\end{equation}
where $P$ is the fluid pressure. So, the fluid velocity \textbf{p} satisfies the same equation as a material surface element moving along the fluid (Batchelor \cite{Batchelor}). (\ref{eq1}) may be viewed to constitute a \textit{Weber Transform} (Lamb \cite{Lamb}, Kuznetsov \cite{Kuznetsov}).

The fluid impulse velocity \textbf{p} gives the total momentum change required at each point to generate from rest a particular fluid motion at that point. Buttke \cite{Buttke}, \cite{ButtkeChorin} and Russo and Smereka \cite{RussoSmereka} showed that \textbf{p}, as defined in (\ref{eq1}) with the geometric gauge (\ref{eq2}), can be interpreted as the fluid impulse density per unit mass. 

The extension of the fluid impulse formulation for a compressible fluid was considered by Tur and Yanovsky \cite{TurYanovsky}, Buttke \cite{ButtkeChorin}, and Shivamoggi \cite{Shivamoggi}. A suitable expression for the fluid impulse was given by Shivamoggi \cite{Shivamoggi} which affords the development of an appropriate impulse formulation for compressible Euler equations.

In this paper, we consider the Kuz'min-Oseledets formulations for the fluid impulse density for incompressible and compressible barotropic Euler equations. We give exact solutions for the fluid impulse density and discuss their physical significance.

\section{Exact Solutions for Incompressible Flows}\par
We will construct exact solutions of the impulse equation (\ref{eq4}) for the incompressible case in a cylindrical geometry and generalize a result by Russo and Smereka \cite{RussoSmereka}. 

\begin{flushleft}
\textbf{(i) Purely Swirling Flow} 
\end{flushleft}
 \par Consider a steady flow with cylindrical symmetry, and vorticity given by
\begin{equation}
\omega = f(r) \hat{z} = \left< 0,0,f(r) \right>.
\label{eq5}
\end{equation}
In order to maintain cylindrical symmetry, the velocity field must show no $\theta$ dependence, 

\begin{equation}
\textbf{v} = \left< v_r (r,z),v_\theta (r,z), v_z (r,z) \right>.
\label{eq7}
\end{equation}
Noting that

\begin{equation}
\omega = \nabla \times \textbf{v} = \left< \frac{1}{r} \frac{\partial v_z}{\partial \theta}-\frac{\partial v_\theta}{\partial z}, \frac{\partial v_r}{\partial z}-\frac{\partial v_z}{\partial r}, \frac{1}{r} \frac{\partial r v_\theta}{\partial r}-\frac{1}{r} \frac{\partial v_r}{\partial \theta} \right> = \langle \omega_r, \omega_\theta, \omega_z \rangle.
\label{eq6}
\end{equation}
we obtain,

\newpage

\begin{equation}
\omega_r = 0 = - \frac{\partial v_\theta}{\partial z}
\label{eq8}
\end{equation}

\begin{equation}
\omega_\theta = 0 = \frac{\partial v_r}{\partial z}-\frac{\partial v_z}{\partial r}
\label{eq9}
\end{equation}

\begin{equation}
\omega_z = f(r) = \frac{1}{r} \frac{\partial r v_\theta}{\partial r}.
\label{eq10}
\end{equation}
Equation (\ref{eq8}) implies that $v_\theta$ is a function of $r$ only. This implies in turn that (\ref{eq10}) is an ordinary differential equation which can be integrated to give

\begin{equation}
v_\theta(r) = \frac{1}{r} \int_{0}^{r} sf(s)ds.
\label{eq11}
\end{equation}

If we further assume that the flow is effectively two dimensional (i.e., it is planar and has no variation in the $z$-direction) and incompressible, then we have
\begin{equation}
v_r(r) = \frac{c}{r}
\label{eq12}
\end{equation}

\begin{equation}
v_z(r) = 0,
\label{eq14}
\end{equation}
$c$ being an arbitrary constant.

The velocity field (\ref{eq11})-(\ref{eq14}) contains a singularity at the origin. If the region occupied by the fluid includes the origin, then we have to choose $c=0$ to remove the singularity. (\ref{eq12}) then leads to

\begin{equation}
v_r(r) = 0.
\label{eq15}
\end{equation}

Now, let us use (\ref{eq11}), (\ref{eq14}) and (\ref{eq15}) in the fluid impulse equation (\ref{eq4}) with the geometric gauge, which in cylindrical coordinates yields,

\begin{equation}
\frac{\partial p_r}{\partial t} + \frac{\partial p_r}{\partial r}v_r+\frac{1}{r}\frac{\partial p_r}{\partial \theta}v_\theta+\frac{\partial p_r}{\partial z}v_z-\frac{p_\theta}{r}v_\theta = -\frac{\partial v_r}{\partial r} p_r - \frac{\partial v_\theta}{\partial r} p_\theta - \frac{\partial v_z}{\partial r}p_z
\label{eq18}
\end{equation}

\begin{multline}
\frac{\partial p_\theta}{\partial t}+\frac{\partial p_\theta}{\partial r}v_r +\frac{1}{r} \frac{\partial p_\theta}{\partial \theta}v_\theta+\frac{\partial p_\theta}{\partial z}v_z+\frac{p_r}{r}v_\theta = -\frac{1}{r}\frac{\partial v_r}{\partial \theta}p_r+\frac{v_\theta}{r}p_r-\frac{1}{r}\frac{\partial v_\theta}{\partial \theta}p_\theta\\
-\frac{v_r}{r}p_\theta-\frac{1}{r} \frac{\partial v_z}{\partial \theta}p_z
\label{eq19}
\end{multline}

\begin{equation}
\frac{\partial p_z}{\partial t}+\frac{\partial p_z}{\partial r}v_r+\frac{1}{r}\frac{\partial p_z}{\partial \theta}v_\theta+\frac{\partial p_z}{\partial z}v_z = -\frac{\partial v_r}{\partial z}p_r-\frac{\partial v_\theta}{\partial z}p_\theta-\frac{\partial v_z}{\partial z}p_z.
\label{eq20}
\end{equation}

Rewriting the velocity field as,

\begin{equation}
\textbf{v}=\left< 0, \frac{1}{r} \int_{0}^{r} s f(s) ds,0 \right> \equiv \left<0,U(r),0 \right>,
\label{eq21}
\end{equation}

\noindent equations (\ref{eq18})-(\ref{eq20}) give:

\begin{equation}
\frac{\partial p_r}{\partial t}+\frac{1}{r} \frac{\partial p_r}{\partial \theta} U-\frac{p_\theta}{r}U = -U' p_\theta
\label{eq22}
\end{equation}

\begin{equation}
\frac{\partial p_\theta}{\partial t}+\frac{1}{r} \frac{\partial p_\theta}{\partial \theta}U=0
\label{eq23}
\end{equation}

\begin{equation}
\frac{\partial p_z}{\partial t}+\frac{1}{r}\frac{\partial p_z}{\partial \theta}U=0.
\label{eq24}
\end{equation}

Thanks to cylindrical symmetry, equations (\ref{eq22})-(\ref{eq24}) become

\begin{equation}
\frac{\partial p_r}{\partial t}-\frac{p_\theta}{r}U=-U' p_\theta
\label{eq25}
\end{equation}

\begin{equation}
\frac{\partial p_\theta}{\partial t}=0
\label{eq26}
\end{equation}

\begin{equation}
\frac{\partial p_z}{\partial t} = 0.
\label{eq27}
\end{equation}

Equations (\ref{eq26}) and (\ref{eq27}) imply that $p_\theta$ and $p_z$ must be unknown functions of $r$ and $z$ only. Since $p_\theta$ is time independent, equation (\ref{eq25}) may be integrated with respect to $t$ to give

\begin{equation}
p_r(r,z,t) = -U' p_\theta(r,z)t+U \frac{p_\theta}{r}t = -r p_\theta \bigg(\frac{U}{r}\bigg) 't.
\label{eq28}
\end{equation}
Thus, while the azimuthal and axial components of the fluid impulse density are time independent, the radial component grows linearly in time. This gives the general solution,

\begin{equation}
\textbf{p}(r,z,t)= \left< -r \bigg( \frac{U}{r} \bigg)' p_\theta(r,z)t,p_\theta(r,z),p_z(r,z) \right>,
\label{eq29}
\end{equation}
where $p_\theta$ and $p_z$ are determined by initial conditions.

The Russo-Smereka \cite{RussoSmereka} solution corresponds to taking,

\begin{equation}
p_{\theta}(r,z)=U(r).
\label{27e}
\end{equation}
Using (\ref{27e}), (\ref{eq28}) leads to

\begin{equation}
p_r(r,t)=-r\Big(\frac{U}{r}\Big)'Ut
\label{28e}
\end{equation}
which is in disagreement with the Russo-Smereka \cite{RussoSmereka} solution,

\begin{equation}
p_r(r,t) = -U U't.
\label{29e}
\end{equation}
For the case,

\begin{equation}
U(r)=cr
\label{30e}
\end{equation}
$c$ being a constant, (\ref{28e}) gives

\begin{equation}
p_r(r,t) = 0
\label{31e}
\end{equation}
while the Russo-Smereka \cite{RussoSmereka} solution (\ref{29e}) gives

\begin{equation}
p_r(r,t)=-c^2rt.
\label{32e}
\end{equation}

\begin{flushleft}
\textbf{(ii) Purely Axial Flow} 
\end{flushleft}

Next, consider the cylindrically symmetrical axial flow, 

\begin{equation}
\textbf{v}=f(r) \hat{z} = \left<0,0,f(r)\right>
\label{eq30}
\end{equation}
The vorticity for this flow is given by,

\begin{equation}
\omega = \nabla \times \textbf{v} = \left<0,-f',0\right>.
\label{eq31}
\end{equation}
Upon substitution into equation (\ref{eq4}), we obtain

\begin{equation}
\frac{\partial p_r}{\partial t} + f \frac{\partial p_r}{\partial z}=-\frac{\partial f}{\partial r}p_z=-f' p_z\linebreak
\label{eq32}
\end{equation}
\begin{equation}
\frac{\partial p_\theta}{\partial t}+f \frac{\partial p_\theta}{\partial z}=-\frac{\partial f}{\partial \theta}p_z = 0
\label{eq33}
\end{equation}
\begin{equation}
\frac{\partial p_z}{\partial t}+f\frac{\partial p_z}{\partial z}=-\frac{\partial f}{\partial z}p_z=0
\label{eq34}
\end{equation}

Since the flow is translation invariant in the axial direction, it makes sense to choose an impulse density exhibiting the same property. Equations (\ref{eq32})-(\ref{eq34}) then become 

\begin{equation}
\frac{\partial p_r}{\partial t}=-f' p_z
\label{eq35}
\end{equation}

\begin{equation}
\frac{\partial p_\theta}{\partial t}=0
\label{eq36}
\end{equation}

\begin{equation}
\frac{\partial p_z}{\partial t}=0
\label{eq37}
\end{equation}

The azimuthal and axial components of the impulse density are time independent, allowing us to integrate equation (\ref{eq35}) in time giving:

\begin{equation}
\textbf{p}(r,t)=\left< -f' p_z(r)t,p_\theta(r),p_z(r) \right>
\label{eq38}
\end{equation}
Thus, while the azimuthal and axial components of the impulse density remain constant in time the radial component again grows linearly with time.

\section{Exact Solution for Compressible, Barotropic Flows}

We now consider a compressible, barotropic fluid flow. Assuming that the flow field shows cylindrical symmetry, we consider the velocity field,

\begin{equation}
\textbf{v}=\left<V,U(r),0\right>
\label{eq39}
\end{equation}
In the limit $V \Rightarrow 0$ , this reduces to the velocity profile (\ref{eq21}) considered in the incompressible case. We will see in the following that $V$ can actually be taken to be the compressibility parameter.

For a compressible fluid, the mass conservation equation in cylindrical coordinates is

\begin{equation}
\frac{\partial \rho}{\partial t}+\frac{1}{r}\frac{\partial }{\partial r}(r \rho v_r)+\frac{1}{r}\frac{\partial}{\partial \theta}(\rho v_\theta)+\frac{\partial }{\partial z}(\rho v_z) = 0.
\label{eq40}
\end{equation}
Substituting the velocity field (\ref{eq39}), equation (\ref{eq40}) becomes

\begin{equation}
\frac{\partial \rho}{\partial t}+\frac{V}{r}\frac{\partial }{\partial r}(r \rho)=0
\label{eq41}
\end{equation}
which may be rewritten as

\begin{equation}
\frac{\partial }{\partial t}(r \rho)+V \frac{\partial }{\partial r}(r \rho)=0.
\label{eq42}
\end{equation}
Putting,

\begin{equation}
\bar{\rho} \equiv r \rho,
\label{eq43}
\end{equation}
equation (\ref{eq42}) becomes

\begin{equation}
\frac{\partial }{\partial t}(\bar{\rho}) + V \frac{\partial }{\partial r}(\bar{\rho})=0,
\label{eq44}
\end{equation}
which has a traveling wave solution,

\begin{equation}
\bar{\rho}(r,t) = f(r-Vt).
\label{eq45}
\end{equation}
The function $f(r,t)$ is determined by boundary conditions and should be such that $\rho$ that is positive definite. Using (\ref{eq45}), (\ref{eq43}) gives

\begin{equation}
\rho(r,t)=\frac{1}{r}f(r-Vt).
\label{eq46}
\end{equation}

Several observations pertaining to (\ref{eq46}) are now in order. Note that the variations in density move outwards from the center like a traveling wave with propagation speed $V$ and its magnitude falls off as $r^{-1}$. So, this solution may be considered to be a density wave riding on a steady Burgers \footnote{Burgers vortex describes the interplay between the intensification of vorticity due to the imposed straining flow and the diffusion of vorticity due to the action of viscosity.} type vortex flow. In the limit $V \Rightarrow 0,$ the density becomes time-independent. Next, we observe that the divergence of the velocity field (\ref{eq39}) in cylindrical coordinates is

\begin{equation}
\nabla \cdot \textbf{v} = \frac{1}{r} \frac{\partial }{\partial r}(r v_r)+\frac{1}{r} \frac{\partial v_\theta}{\partial \theta}+\frac{\partial v_z}{\partial z} = \frac{V}{r}.
\label{eq47}
\end{equation}
Barring the point $r=0,$ in the limit $V \Rightarrow 0,$ (\ref{eq47}) leads to

\begin{equation}
\nabla \cdot \textbf{v} = 0
\label{eq48}
\end{equation}
confirming the velocity field is solenoidal in the limit $V \Rightarrow 0.$ Thus, $V$ can indeed be taken to be a compressibility parameter characterizing the non-solenoidal aspects of $\textbf{v}.$ Note that, in the limit $V \Rightarrow 0,$ we may choose $f(r)=C r,$ which makes the density constant as is necessary for an incompressible fluid.

Let us next consider the components of the fluid impulse density. Equations (\ref{eq18})-(\ref{eq20}) for the evolution of the cylindrical components of the impulse density (which remain applicable to a compressible fluid, under the barotropic assumption), on substituting the velocity field (\ref{eq39}), become

\begin{equation}
\frac{\partial p_r}{\partial t} + \frac{\partial p_r}{\partial r}V+\frac{1}{r} \frac{\partial p_r}{\partial \theta}U-\frac{p_\theta}{r}U=-U'p_\theta
\label{eq49}
\end{equation}

\begin{equation}
\frac{\partial p_\theta}{\partial t} + \frac{\partial p_\theta}{\partial r}V+\frac{1}{r}\frac{\partial p_\theta}{\partial \theta}U+\frac{p_r}{r}U=\frac{U}{r}p_r-\frac{V}{r}p_\theta
\label{eq50}
\end{equation}

\begin{equation}
\frac{\partial p_z}{\partial t}+\frac{\partial p_z}{\partial r}V+\frac{1}{r} \frac{\partial p_z}{\partial \theta}U=0
\label{eq51}
\end{equation}

\noindent Assuming again that the flow is cylindrically symmetrical, equations (\ref{eq49})-(\ref{eq51}) become

\begin{equation}
\frac{\partial p_r}{\partial t}+\frac{\partial p_r}{\partial r}V-\frac{p_\theta}{r}U=-U' p_\theta
\label{eq52}
\end{equation}

\begin{equation}
\frac{\partial p_\theta}{\partial t}+ \frac{\partial p_\theta}{\partial r}V=-\frac{V}{r}p_\theta
\label{eq53}
\end{equation}

\begin{equation}
\frac{\partial p_z}{\partial t}+\frac{\partial p_z}{\partial r}V=0
\label{eq54}
\end{equation}

Equation (\ref{eq54}) has a traveling wave solution,

\begin{equation}
p_z(r,t)=g_z(r-Vt)
\label{eq55}
\end{equation}

Next, note that equation (\ref{eq53}) may be rewritten as

\begin{equation}
\frac{\partial (r p_\theta)}{\partial t}+V \frac{\partial (r p_\theta)}{\partial r}=0,
\label{eq56}
\end{equation}
and also has a traveling wave solution,

\begin{equation}
p_\theta(r,t)=\frac{1}{r}g_\theta(r-Vt).
\label{eq57}
\end{equation}

Finally, equation (\ref{eq52}) may be rewritten as

\begin{equation}
\frac{\partial p_r}{\partial t}+V \frac{\partial p_r}{\partial r}= \frac{p_\theta}{r}U-U'p_\theta=-r p_\theta \left( \frac{U}{r} \right)'
\label{eq58}
\end{equation}
the solution of which is

\begin{equation}
p_r(r,t) = -g_\theta(\xi)\frac{\partial }{\partial \xi} \int_{0}^{t}\frac{U(\xi+V  \tau)}{\xi+V \tau}d\tau
\label{eq59}
\end{equation}
where,

\begin{equation}
\xi \equiv r-V t.
\label{eq60}
\end{equation}
(\ref{eq59}) may be rewritten as,

\begin{equation}
p_r(r,t) = r p_\theta(r-Vt)\frac{\partial }{\partial r} \int_{0}^{t} \frac{U(r-V(t-\tau))}{r-V(t-\tau)}d\tau
\label{eq61}
\end{equation}
(\ref{eq61}) expresses the plausible fact that the solution at distance $r$ depends on the source element not only at $r$, but on all the source elements at a distance less than $r$ by the distance required to propagate the effect (with velocity $V$) from an earlier time $\tau$ to $t$. In the limit $V\Rightarrow 0,$ (\ref{eq61}) reduces to the corresponding result (\ref{eq28}) for the incompressible case. On the other hand, in the extreme compressible limit $V \Rightarrow \infty,$ (\ref{eq61}) leads to $p_r \Rightarrow 0.$

\section{Discussion}
In this paper, we have considered the Kuz'min-Oseledets formulations for the fluid impulse density for incompressible and compressible barotropic Euler equations. We have given exact solutions for the fluid impulse density and discussed their physical significance. The compressible barotropic exact solution describes a density wave riding on a steady Burgers type vortex flow with the fluid impulse velocity component in the wave-propagation direction growing monotonically with time.

\section*{Acknowledgements}
BKS is thankful to Dr. S. Kurien for helpful discussions on impulse formulations.

\end{document}